# Edge State and Intrinsic Hole Doping in Bilayer Phosphorene

Toshihito Osada*

*Institute for Solid State Physics, University of Tokyo,*

*5-1-5 Kashiwanoha, Kashiwa, Chiba 277-8581, Japan.*

Using a simple LCAO model by Harrison, we have qualitatively studied the edge state of bilayer phosphorene, which is a unit structure of the layered crystal of black phosphorus. This model successfully reproduces the isolated edge state in the bulk gap in monolayer phosphorene. In bilayer phosphorene, however, it shows that edge states are almost buried in the valence band and there is no isolated midgap edge state at the zigzag edge. Since the buried edge state works as acceptor, holes are doped from the edge state into the bulk. This gives a possible explanation for p-type conduction in undoped black phosphorus. Under the vertical electric field, the intrinsic hole doping is reduced because a part of edge states move into the gap. These features of bilayer phosphorene might be better suited for device application.

Recently, phosphorene, an atomic layer of phosphorus, was realized by the mechanical exfoliation from the layered bulk crystal of black phosphorus [1]. It is the third single component two-dimensional (2D) crystal following graphene and silicene. In contrast to graphene, which is a 2D zerogap conductor, phosphorene is a 2D semiconductor with a finite band gap, so that it can be easily used as a conventional device material. Moreover, high hole mobility, 286 $cm^2$/Vs at room temperature, was reported in phosphorene [1]. So, phosphorene is a promising p-type 2D semiconductor complementary to the n-type atomic layers like $MoS_2$ [1, 2]. In fact, its FET characteristics has been extensively studied mainly in multilayer phosphorene [3-7].

The electronic structure of phosphorene was originally discussed in 1981 based on the tight-binding calculation [8] in order to study the electronic structure of black phosphorus [9, 10]. After the recent rediscovery of phosphorene, many studies have been performed on its electronic structure based on the first-principle calculation [11-14] and the tight-binding approach [15]. They concluded that phosphorene is a 2D semiconductor with a direct gap, much larger than bulk black phosphorus. In addition, the edge state of monolayer phosphorene has been studied by considering phosphorene nanoribbon based on first-principle [16] and tight-binding methods [17]. At the pristine zigzag edge, a half-filled edge state with one-dimensional dispersion appears in the middle of the bulk energy gap. At the armchair edge, a pair of edge states are formed in the gap. Since the Fermi energy exists between them, the system becomes semiconducting.

The layered crystal of black phosphorus is formed by alternative stacking of monolayer phosphorene, so that its unit structure is bilayer phosphoren. Therefore, it is expected that basic properties of black phosphorus relates to those of bilayer. It is known that undoped black phosphorus is a p-type semiconductor [10], but the mechanism of natural p-type doping has not been clarified yet. As for the electronic structure of bilayer, the band structure has already been discussed [15, 18], whereas the edge state feature has not been clarified yet.

In this paper, we qualitatively discuss the electronic structure of bilayer phosphorene, especially its possible edge states, using a simple LCAO model proposed by Harrison [19]. In this model, we assume the matrix element (transfer integral) between adjacent two atoms as $V_{ll'm}(d) = \eta_{ll'm}\hbar^2/m_0 d^2$, where $d$ and $m_0$ are the inter-atomic distance and the electron rest mass, respectively. $l$ and $l'$ denote the orbital azimuthal quantum numbers ($s$, $p$) of two atoms, and $m$ is the common orbital magnetic quantum number ($\sigma$, $\pi$) along the axis connecting two atoms. Although the dimensionless parameter $\eta_{ll'm}$ should be chosen depending on the crystal structure, we simply adopt the adjusted values $\eta_{ss\sigma} = -1.40$, $\eta_{sp\sigma} = 1.84$, $\eta_{pp\sigma} = 3.24$, and $\eta_{pp\pi} = -0.81$ given by Harrison [19]. Although we cannot expect quantitative accuracy in this model, we can easily extract qualitative features of electronic structure.

First, we apply the present model to monolayer phosphorene to check its applicability. The crystal structure of monolayer phosphorene is illustrated in Fig. 1(a). It forms a puckered layer, in which phosphorus atoms are located on two parallel planes. The potential difference $2\Delta$ can be introduced between two planes by applying the electric field normal to the layer. Fig. 1(b) is the projection of phosphorene crystal onto the 2D plane. The dashed rectangle indicates a unit cell, which contains four atoms labeled A, B, A', and B'. Their positions in the unit cell are $\tau_A = (uc, 0, vb)$, $\tau_B = ((1/2 - u)c, a/2, vb)$, $\tau_{A'} = -\tau_A$, and $\tau_{B'} = -\tau_B$, where $c$, $a$, and $b$ are lattice constants, and $u$ and $v$ are dimensionless parameters shown in Fig. 1(a) and (b). We use the values of black phosphorus; $a = 3.314$ Å, $b = 10.478$ Å, $c = 4.376$ Å, $u = 0.08056$, and $v = 0.10168$ [8]. The lattice displacement vectors are defined as $\boldsymbol{d}^{(1)} = \boldsymbol{\tau}_B - \boldsymbol{\tau}_A$, $\boldsymbol{d}^{(1)\prime} = (\boldsymbol{\tau}_B - \boldsymbol{a}) - \boldsymbol{\tau}_A$, $\boldsymbol{d}^{(2)} = \boldsymbol{\tau}_{A\prime} - \boldsymbol{\tau}_A$, and $\bar{\boldsymbol{d}}^{(2)} = \boldsymbol{\tau}_{B\prime} - \boldsymbol{\tau}_B$ with $\boldsymbol{a} = (0, a, 0)$ and $\boldsymbol{c} = (c, 0, 0)$.

We take into account up to the second nearest neighbor couplings, since the matrix element $V_{ll\prime m}(d)$ is applicable only to adjacent atoms. The effective Hamiltonian is represented as the $16 \times 16$ matrix with the base $\{A_s(\boldsymbol{k}), A_x(\boldsymbol{k}), A_y(\boldsymbol{k}), A_z(\boldsymbol{k}), B_s(\boldsymbol{k}), B_x(\boldsymbol{k}), B_y(\boldsymbol{k}), B_z(\boldsymbol{k}), A'_s(\boldsymbol{k}), A'_x(\boldsymbol{k}), A'_y(\boldsymbol{k}), A'_z(\boldsymbol{k}), B'_s(\boldsymbol{k}), B'_x(\boldsymbol{k}), B'_y(\boldsymbol{k}), B'_z(\boldsymbol{k})\}$, where $A_s(\boldsymbol{k})$ denotes the Bloch sum of the wave functions of the $3s$ orbital in A sites, $B_x(\boldsymbol{k})$ denotes that of $3p_x$ orbital in B sites, and so on. The 2D wave number is represented by $\boldsymbol{k} = (k_x, k_y, 0)$. The Hamiltonian is written as the following block matrix, of which elements ($M_0$, $M_1$, $M_2^{\pm}$, etc.) are $4 \times 4$ matrices.

$$H_{\text{mono}}(\boldsymbol{k}; \Delta) = \begin{pmatrix} M_0 + \Delta & M_1 & M_2^+ & 0 \\ {}^tM_1^* & M_0 + \Delta & 0 & M_2^- \\ {}^tM_2^{+*} & 0 & M_0 - \Delta & {}^tM_1^* \\ 0 & {}^tM_2^{-*} & M_1 & M_0 - \Delta \end{pmatrix}. \quad (1)$$

The diagonal $4 \times 4$ matrices express the energies of four atomic cites with

$$M_0 = \begin{pmatrix} \varepsilon_s & 0 & 0 & 0 \\ 0 & \varepsilon_p & 0 & 0 \\ 0 & 0 & \varepsilon_p & 0 \\ 0 & 0 & 0 & \varepsilon_p \end{pmatrix}. \quad (2)$$

Here, $\varepsilon_s = -17.10$ eV and $\varepsilon_p = -8.33$ eV are the energy levels of $3s$ and $3p$ orbitals of phosphorus, respectively [19]. They are shifted by $\pm\Delta$ due to the external vertical electric field. Note that $\pm\Delta$ is multiplied by the $4 \times 4$ unit matrix. The nearest neighbor coupling between A and B (A' and B') is represented as the following.

$$M_1 = \begin{pmatrix} E_{ss}^{(1)} g_1^+ & E_{sx}^{(1)} g_1^+ & E_{sy}^{(1)} g_1^- & 0 \\ -E_{sx}^{(1)} g_1^+ & E_{xx}^{(1)} g_1^+ & E_{xy}^{(1)} g_1^- & 0 \\ -E_{sy}^{(1)} g_1^- & E_{xy}^{(1)} g_1^- & E_{yy}^{(1)} g_1^+ & 0 \\ 0 & 0 & 0 & E_{zz}^{(1)} g_1^+ \end{pmatrix}. \tag{3}$$

In the same way, the next nearest neighbor coupling between A and A', and that between B and B' are represented by

$$M_2^\pm = \begin{pmatrix} E_{ss}^{(2)} g_2^\pm & \pm E_{sx}^{(2)} g_2^\pm & 0 & E_{sz}^{(2)} g_2^\pm \\ \mp E_{sx}^{(2)} g_2^\pm & E_{xx}^{(2)} g_2^\pm & 0 & \pm E_{sz}^{(2)} g_2^\pm \\ 0 & 0 & E_{yy}^{(2)} g_2^\pm & 0 \\ -E_{sz}^{(2)} g_2^\pm & \pm E_{xz}^{(2)} g_2^\pm & 0 & E_{zz}^{(2)} g_2^\pm \end{pmatrix}. \tag{4}$$

In $M_1$ and $M_2^\pm$, $E_{ss}^{(i)} = V_{ss\sigma}(d^{(i)})$, $E_{s\alpha}^{(i)} = (d_\alpha^{(i)}/d^{(i)})V_{sp\sigma}(d^{(i)})$, and $E_{\alpha\beta}^{(i)} = \left(d_\alpha^{(i)} d_\beta^{(i)}/{d^{(i)}}^2\right) V_{pp\sigma}(d^{(i)}) + \left(\delta_{\alpha\beta} - d_\alpha^{(i)} d_\beta^{(i)}/{d^{(i)}}^2\right) V_{pp\pi}(d^{(i)})$ are inter-atomic matrix elements, where $\boldsymbol{d}^{(i)} = \left(d_x^{(i)}, d_y^{(i)}, d_z^{(i)}\right)$ and $d^{(i)} = |\boldsymbol{d}^{(i)}|$ ($i = 1, 2$ and $\alpha, \beta = x, y, z$). The phase factors are defined by $g_1^+ = \exp(i\boldsymbol{k} \cdot \boldsymbol{d}^{(1)}) + \exp(i\boldsymbol{k} \cdot \boldsymbol{d}^{(1)\prime})$, $g_1^- = \exp(i\boldsymbol{k} \cdot \boldsymbol{d}^{(1)}) - \exp(i\boldsymbol{k} \cdot \boldsymbol{d}^{(1)\prime})$, $g_2^+ = \exp(i\boldsymbol{k} \cdot \boldsymbol{d}^{(2)})$, and $g_2^- = \exp(i\boldsymbol{k} \cdot \overline{\boldsymbol{d}}^{(2)})$.

By diagonalizing $H_{\text{mono}}(\boldsymbol{k}; \Delta)$, we can obtain the band dispersion of monolayer phosphorene under zero electric field ($\Delta = 0$) as shown in Fig. 1(c). This dispersion qualitatively reproduces the previous tight-binding calculation [8]. A large direct energy gap opens at the Γ point ($\boldsymbol{k} = \boldsymbol{0}$) between the 10-th and 11-th band, and the Fermi level is located in this gap. We can see that sixteen bands are divided into eight pairs. The lowest two pairs, three pairs just below the gap, and three pairs above the gap can be roughly regarded as the $3s$ bands, the $3p$ bonding bands, and the $3p$ anti-bonding bands, respectively. Each pair degenerate at the zone boundary (XM and YM line). This gapless feature reflects the fact that A and A' (B and B') atoms are equivalent in the projected 2D lattice under zero field ($\Delta = 0$) and the effective unit cell is one half of the true unit cell. Once the electric field becomes finite ($\Delta \neq 0$), a gap opens along the XM boundary. At several points, the neighboring band pairs connect with each other forming the Dirac cones with the linear dispersion. This feature might result from the fact that the projected 2D lattice forms a modified honeycomb lattice similar to graphene.

In graphene, the zero-mode edge state appears along the zigzag edge [20, 21]. To study the possible edge state in phosphorene, we have calculated the electronic structure of the phosphorene nanoribbon with finite width based on the above LCAO model. Fig. 2(a) shows the energy dispersion of the nanoribbon parallel to the $y(\boldsymbol{a})$ axis with the "zigzag" edge shown in Fig. 1(b). The width of the nanoribbon was taken as 40 unit cells, and the vertical electric field was set to be zero ($\Delta = 0$). An isolated edge state appears with one-dimensional dispersion in the middle of the main gap. It has two-fold degeneracy corresponding to two sides of the nanoribbon. Since this midgap edge state is stoichiometrically half-filling, it can contribute to metallic conduction at high temperatures where the localization effect is ignored. Under the electric field ($\Delta \neq 0$), the midgap edge state shows the splitting since edge states at both sides of the nanoribbon lie on different planes as seen in Fig. 1(b). We can also see that the bulk gap is slightly reduced by the vertical electric field.

On the other hand, Fig. 2(c) shows the dispersion of the nanoribbon parallel to the $x(\boldsymbol{c})$ axis with the "armchair" edge at zero field. Two edge states are formed in the gap, and each of them is doubly degenerated corresponding to both sides of the nanoribbon. Since the Fermi level exists between them, the whole system is still semiconducting. The gap between two edge states weakly increases with the electric field as shown in Fig. 2(d). This is reasonable because those edge states correspond to the bonding and anti-bonding states of the top and bottom planes at the armchair edge.

The above qualitative features of monolayer agree well with preceding works, although there exist quantitative discrepancies. It means that the present LCAO model is useful for qualitative consideration. So, we apply the present model to bilayer phosphorene. The crystal structure of bilayer and its projection on the $xy$ plane are illustrated in Fig. 3(a) and (b). Two monolayers stack with the shift by $a/2$ in the $y$ direction. Under the electric field, the potential at four atomic planes are expected to be $\pm\Delta'(1 \pm 4v)$, where $2\Delta'$ is the average potential difference between two monolayers. A unit cell contains eight atoms labeled A, B, A', B', C, D, C' and D'. In addition to the nearest neighbor and next nearest neighbor in-plane couplings, we take into account the nearest neighbor interlayer coupling between A' and C (B' and D). Their displacement vectors are $\boldsymbol{d}^{(3)} = (2uc, a/2, 2vb - b/2)$, $\boldsymbol{d}^{(3)\prime} = (2uc, -a/2, 2vb - b/2)$, $\overline{\boldsymbol{d}}^{(3)} = (-2uc, a/2, 2vb - b/2)$, and $\overline{\boldsymbol{d}}^{(3)\prime} = (-2uc, -a/2, 2vb - b/2)$ indicated in Fig. 3(b). The effective Hamiltonian is given by the following $32 \times 32$ matrix

$$H_{\text{bi}}(\boldsymbol{k}; \Delta') = \begin{pmatrix} H_{\text{mono}}(\boldsymbol{k}; 4v\Delta') + \Delta' & H_\perp(\boldsymbol{k}) \\ {}^tH_\perp(\boldsymbol{k})^* & H_{\text{mono}}(\boldsymbol{k}; 4v\Delta') - \Delta' \end{pmatrix}. \tag{5}$$

Here, the interlayer coupling is represented by the $16 \times 16$ matrix,

$$H_\perp(\boldsymbol{k}) = \begin{pmatrix} 0 & 0 & 0 & 0 \\ 0 & 0 & 0 & 0 \\ M_3^+ & 0 & 0 & 0 \\ 0 & {}^tM_3^{-*} & 0 & 0 \end{pmatrix}, \tag{6}$$

where

$$M_3^\pm = \begin{pmatrix} E_{ss}^{(3)} g_3^+ & E_{sx}^{(3)} g_3^+ & E_{sy}^{(3)} g_3^- & \pm E_{sz}^{(3)} g_3^+ \\ -E_{sx}^{(3)} g_3^+ & E_{xx}^{(3)} g_3^+ & E_{xy}^{(3)} g_3^- & \pm E_{xz}^{(3)} g_3^+ \\ -E_{sy}^{(3)} g_3^- & E_{xy}^{(3)} g_3^- & E_{yy}^{(3)} g_3^+ & \pm E_{yz}^{(3)} g_3^- \\ \mp E_{sz}^{(3)} g_3^+ & \pm E_{xz}^{(3)} g_3^+ & \pm E_{yz}^{(3)} g_3^- & E_{zz}^{(3)} g_3^+ \end{pmatrix}. \qquad (7)$$

The inter-atomic matrix elements $E_{ss}^{(3)}$, $E_{s\alpha}^{(3)}$, and $E_{\alpha\beta}^{(3)}$ ($\alpha, \beta = x, y, z$) are defined in the same way as the monolayer case, and the phase factors are defined by $g_3^+ = \exp(i\mathbf{k} \cdot \mathbf{d}^{(3)}) + \exp(i\mathbf{k} \cdot \mathbf{d}^{(3)\prime})$ and $g_3^- = \exp(i\mathbf{k} \cdot \mathbf{d}^{(3)}) - \exp(i\mathbf{k} \cdot \mathbf{d}^{(3)\prime})$. Note that $\mathbf{k} \cdot \overline{\mathbf{d}}^{(3)} = -\mathbf{k} \cdot \mathbf{d}^{(3)\prime}$ and $\mathbf{k} \cdot \overline{\mathbf{d}}^{(3)\prime} = -\mathbf{k} \cdot \mathbf{d}^{(3)}$ since $\mathbf{k} = (k_x, k_y, 0)$. Fig. 3(c) shows the calculated band dispersion of bilayer phosphorene at zero electric field ($\Delta' = 0$). As the valence band top is elevated, the gap is remarkably reduced compared with monolayer. Thirty two bands are divided into eight sets in the same way as monolayer, and each set consists of two pairs of bands. Each pair shows two-fold degeneracy at the zone boundary, and each set has four-fold degeneracy at the zone corner (M point). Some neighboring sets connect with each other, forming the Dirac cones.

Using the same LCAO model, we have also calculated the electronic structure of the bilayer phosphorene nanoribbon with finite width of 40 cells. Fig. 4(a) shows the dispersion of the bilayer nanoribbon parallel to the $y(\mathbf{a})$ axis with the "zigzag" edge under zero electric field ($\Delta' = 0$). Two edge subbands, which correspond to bonding and anti-bonding of edge states on two layers, appear around the main gap. They degenerate at the zone boundary ($k_y = \pm\pi/a$). Each subband is doubly degenerated corresponding to two sides of the nanoribbon. The upper subband is almost tangent to the valence band top, whereas the lower one penetrates the valence band. The most part of both edge states overlap with the valence band in energy. If there was no overlap between edge states and the valence band, the half of edge states would be occupied stoichiometrically. Therefore, holes are doped from the upper edge subband into the valence band. In other words, the upper edge state works as acceptor. This is the intrinsic hole doping mechanism without additional impurities. This might be responsible for the natural p-type doping in the bulk black phosphorus crystal.

Under the vertical electric field ($\Delta \neq 0$), the bulk gap is slightly reduced, and the edge states doubly split as seen in Fig. 4(b) and (c). This split reflects the fact that edge states at both sides of the ribbon lie on different atomic planes as illustrated in Fig. 3(a). As the electric field is increased, the upper two split edge states move into the gap. This causes the reduction of the intrinsic hole doping from the upper edge state.

In monolayer phosphorene, the existence of the midgap edge state might be crucial for device application, since it causes current leakage at room temperature. In contrast, the bilayer phosphorene has fewer midgap state. Therefore, we can expect better device characteristics in bilayer phosphorene. The intrinsic hole doping which can be controlled by the electric field has also potential advantage. The present features of bilayer phosphorene could be experimentally confirmed using the dual-gated FET device, in which the Fermi level (carrier density) and the vertical electric field can be controlled independently.

In summary, we have studied the electronic structure of monolayer and bilayer phosphorene qualitatively using a minimal LCAO model by Harrison. They are 2D semiconductors with the direct gap, which becomes smaller under the vertical electric fields. In contrast to the isolated edge state in the gap in monolayer, most of the edge states overlap with the valence band at the zigzag edge in bilayer. This band configuration causes intrinsic hole doping from the edge state to the bulk in bilayer phosphorene. The doping level can be controlled by the vertical electric field. These features of bilayer phosphorene might be better suited for device application.

**Acknowledgements**     The author thanks to Prof. Y. Akahama and Prof. H. Tajima for arousing our interest in phosphorene.  This work is partially supported by a Grant-in-Aid for Scientific Research on Innovative Areas "Science of Atomic Layers" (No. 25107003) from the Ministry of Education, Culture, Sports, Science, and Technology, Japan.

**Figure 1** (Osada)

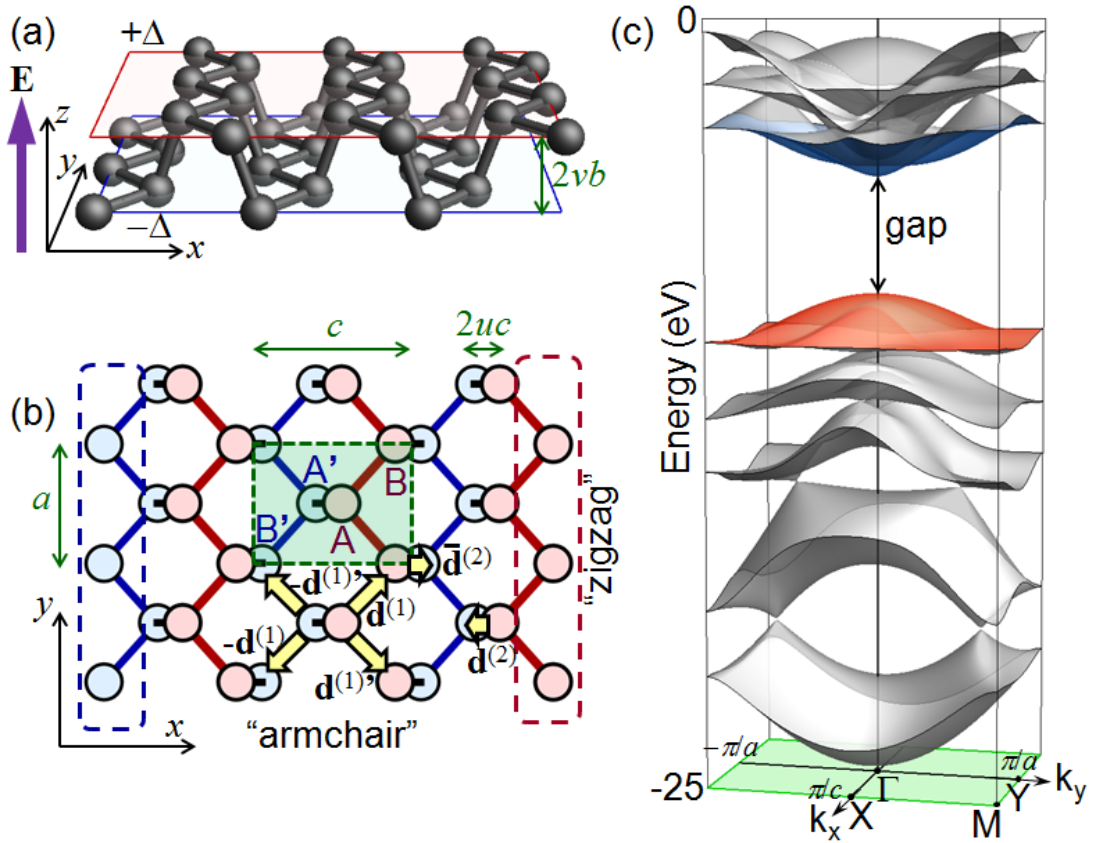

**FIG. 1.** (color online)

(a) Crystal structure of monolayer phosphorene. (b) Projection of phosphorene crystal onto the two-dimensional plane. Dashed rectangle indicates the unit cell, which contains four phosphorus atoms. The "zigzag" and "armchair" edges are also indicated. (c) LCAO band structure of monolayer phosphorene.

**Figure 2** (Osada)

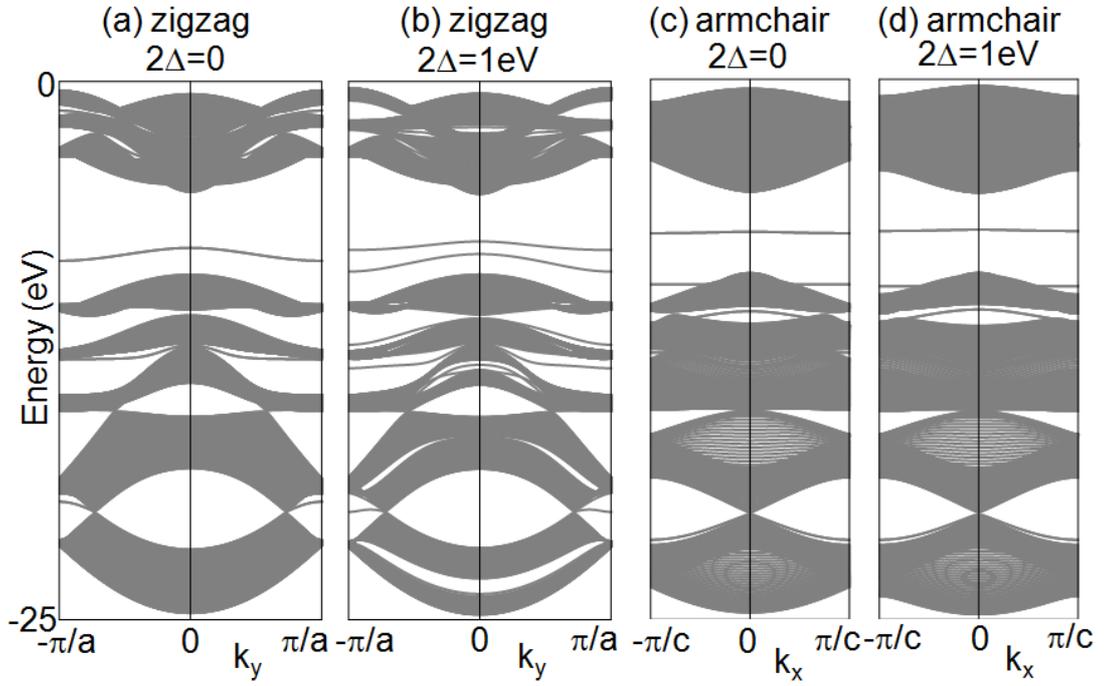

**FIG. 2.**

Energy spectra of monolayer phosphorene nanoribbon. (a) Zigzag edge nanoribbon at zero field ($\Delta = 0$). (b) Zigzag edge nanoribbon under the electric field ($2\Delta = 1$ eV). (b) Armchair edge nanoribbon at zero field ($\Delta = 0$). (d) Armchair edge nanoribbon under the electric field ($2\Delta = 1$ eV).

**Figure 3** (Osada)

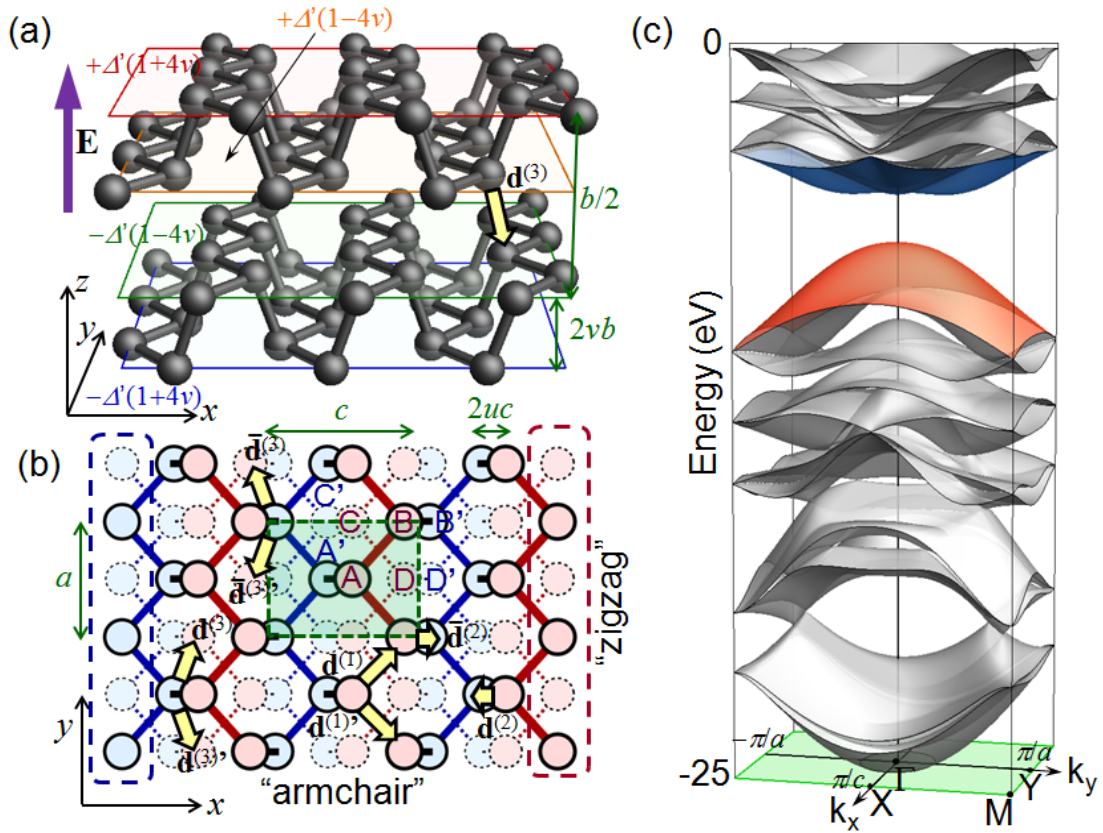

**FIG. 3.** (color online)

(a) Crystal structure of bilayer phosphorene. (b) Projection of bilayer crystal onto the two-dimensional plane. Dashed rectangle indicates the unit cell, which contains eight phosphorus atoms. The "zigzag" and "armchair" edges are also indicated. (c) LCAO band structure of bilayer phosphorene.

**Figure 4** (Osada)

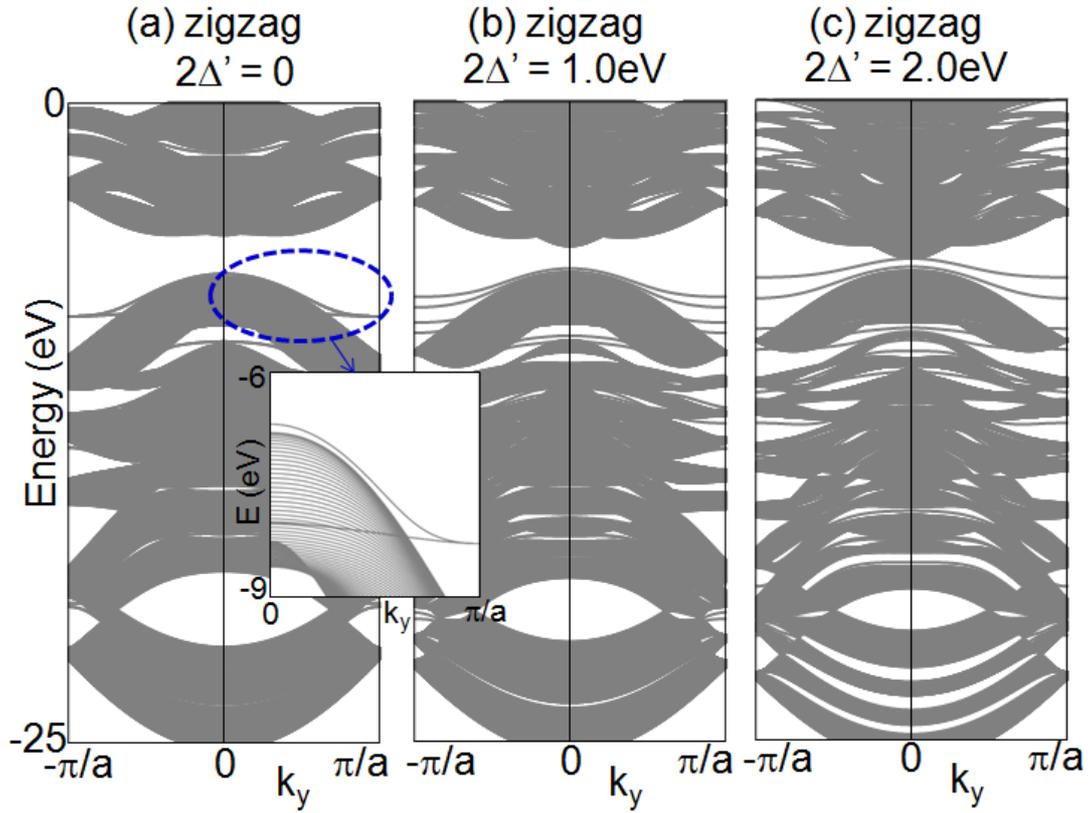

**FIG. 4.** (color online)

Energy spectra of bilayer phosphorene nanoribbon with zigzag edge under the vertical electric field. (a) Zero electric field ($\Delta = 0$). Inset shows the detail of edge states. Upper edge subband works as acceptors. (b) $2\Delta = 1$ eV. (c) $2\Delta = 2$ eV.